\documentclass[%
 reprint,
 amsmath,amssymb,
 aps,super,compress,sort,comma
]{revtex4-2}
\usepackage{xr}
\makeatletter
\newcommand*{\addFileDependency}[1]{
\typeout{(#1)}
\@addtofilelist{#1}
\IfFileExists{#1}{}{\typeout{No file #1.}}
}\makeatother

\newcommand*{\myexternaldocument}[1]{%
\externaldocument{#1}%
\addFileDependency{#1.tex}%
\addFileDependency{#1.aux}%
}
\myexternaldocument{SI}

\usepackage{amsmath}
\usepackage{graphicx}
\usepackage{dcolumn}
\usepackage{bm}
\begin{document}

\preprint{APS/123-QED}
\title{Strain engineering of topological magnons in chromium trihalides from first-principles}
\author{Dorye L. Esteras}
\author{José J. Baldoví}
\email{j.jaime.baldovi@uv.es}
\affiliation{Instituto de Ciencia Molecular, Universitat de València, Catedrático José Beltrán 2, 46980 Paterna, Spain.
}

\begin{abstract}
Recent experiments evidence the direct observation of spin waves in chromium trihalides and a gap at the Dirac points of the magnon dispersion in bulk CrI$_3$. However, the topological origin of this feature remains unclear and its emergence at the 2D limit has not yet been proven experimentally. Herein, we perform a fully self-consistent ab initio analysis that supports the presence of topological magnons in chromium trihalides monolayers. Our results confirm the existence of a gap around the K high-symmetry point in the linear magnon dispersion of CrI$_3$, which originates as a direct consequence of intralayer Dzyaloshinskii-Moriya (DM) interaction. In addition, our orbital resolved analysis reveals the microscopic mechanisms that can be exploited using strain engineering to increase the strength of the DM interaction and thus control the gap size in CrI$_3$. This paves the way to the further development of this family of materials as building-blocks for topological magnonics at the limit of miniaturization.
\end{abstract}

\maketitle

\section{Introduction}  
Two-dimensional (2D) materials have become a focus of intense research due to their unique physical and chemical properties \cite{2dspintronics,cri3,graphene}. The recent discovery of long-range magnetic order in 2D van der Waals layered materials down to the single layer has opened new frontiers for the exploration of 2D magnetism and the development of novel spin-based applications at the nanoscale \cite{efren,feps3}. A stimulating challenge in this direction is the control of collective spin excitations, also known as spin waves –and their quanta magnons–, which are present in magnetically ordered materials \cite{demokritov,buildingblocks,magnonics,magnonics2,roadmap}. Magnons can be used to store, process and transmit information, and thus open the door to low-energy consumption nanodevices with higher tunability, excellent integrability and many other advantages pursued by the emergent field of magnonics \cite{chumak2015magnon,rodin,neusser2009magnonics,nikitovmagnonics}. Furthermore, recent research has led to new insights into the topology of these collective magnetic excitations, indicating that non-trivial topological phases of matter can arise within magnon band structures \cite{topologicalmagnonics,eltonsantos,experimentcrbr3nuevo,experimentcrcl3}. This offers a promising platform to drive topological transitions using magnetic fields and generate topologically protected spin currents. Another well-known advantage of 2D materials is the possibility of tuning their properties via mechanical strain, which has been culminated in the field of straintronics \cite{roy2011hybrid,bukharaev2018straintronics,roldan2015strain,castellanos2013local}. In the context of 2D magnetic materials, strain engineering has allowed the enhancement of the magnetic anisotropy energy and spin-spin interactions of chromium trihalides (CrX$_3$, X = Cl, Br, I) \cite{straincrx3,wustrain,mukherjee2019strain,dmistrainfield}, and has been recently extended to the realm of magnonics in CrSBr \cite{esterascrsbr}. \\ 

Ferromagnetic (FM) honeycomb lattices provide ideal building-blocks to search for topological magnon excitations. The symmetry of these systems creates linear Dirac dispersions in the magnon band structure around the K high-symmetry point, where time-reversal symmetry is broken \cite{diracmagnonshoneycomb,topologicalhoneycombs}.  Consequently, the CrX$_3$ family stays in the line of fire \cite{costa,aguilera,baidya2018tunable,milosevicstackings} since the experimental observation of spin waves in all three compounds has been reported \cite{experimentcrcl3,crbrviejuno,experimentcrbr3,experimentcrbr3nuevo,eltonsantos}.  Moreover, the last empirical evidence points towards the topological nature of these magnons, suggesting that this class of 2D materials can be useful for topological magnonic applications \cite{tunablemagnon,crbr3magnontunelling,crcl3bed,crcl3gigahertz,stroppamagnonictrihalides}. The non-trivial topology arises from gapped Dirac cones in the magnon dispersion, but depends crucially on the origin of this gap, which has been experimentally measured in bulk CrI$_3$ \cite{experimentcri3viejete,eltonsantos}. However, the mechanism behind this feature still remains controversial \cite{olsenpegagap,discussdmikitaev,discusskitaev,discussioninterlayer,confusionkitaevdmi} and the survival of the topological gap at the 2D limit is unexplored. Indeed, the possibility of carrying out inelastic neutron scattering measurements in monolayer chromium trihalides represents a gigantic challenge, which imposes an important restriction to understand the origin of the gap in the absence of interlayer magnetic exchange interactions.\\

To unveil the microscopic origin of topologically protected magnons in 2D chromium trihalides, herein, we perform a fully self-consistent Hubbard-corrected ab initio investigation of CrCl$_3$, CrBr$_3$ and CrI$_3$ followed by a derived Wannier-based tight-binding model. Our optimized self-consistent parameter-free workflow predicts the presence of a clear topological gap in CrI$_3$ and a very small gap of CrBr$_3$ in the absence of interlayer interactions, revealing important details about the underlying mechanisms behind its formation. In particular, we illustrate the crucial role of Dzyaloshinskii-Moriya (DM) interactions and rationalize it by means of an orbital-resolved magnetic exchange analysis. Then, we apply biaxial strain engineering in order to enhance DM interaction, which results in the possibility of observing a drastic increase of the topological gap in CrI$_3$. Our results pave the way for the use of 2D chromium trihalides as building components for topological magnonic devices.

\section{Results and discussion}
\begin{figure*}[h!]
\includegraphics[scale=0.5]{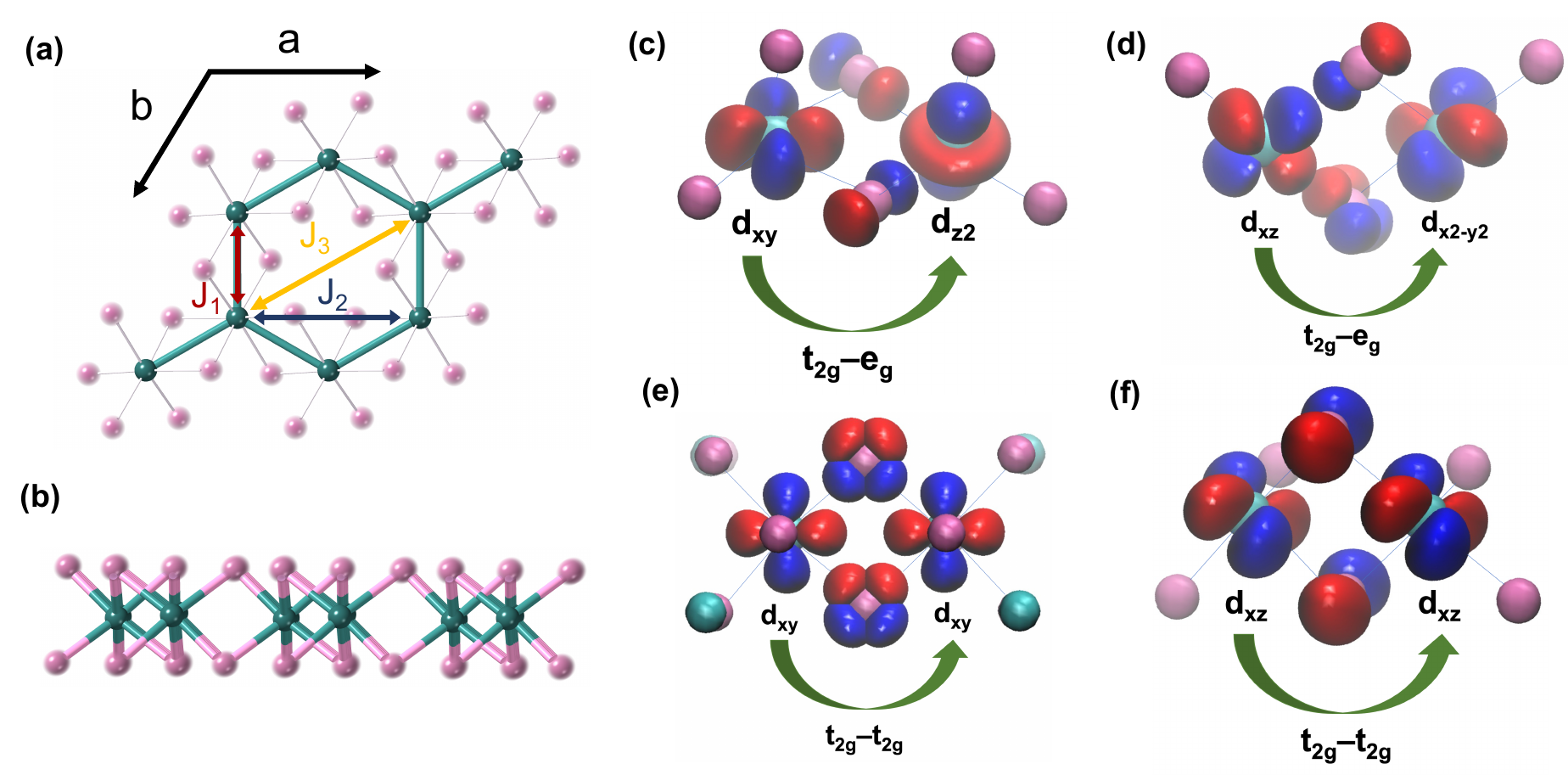}
\caption{ (a) Top view of a CrX$_3$ monolayer indicating the main exchange interactions. (b) Side view of the same structure. (c–f) Calculated maximally localized Wannier functions representing the atomic orbitals responsibles of J$_1$ interaction in CrX$_3$. (c,d) $t_{2g}$-$e_{g}$ (FM) and (e,f) $t_{2g}$-$t_{2g}$ (AFM) pathways.}
\label{fig:1}
\end{figure*}

At low temperature, chromium trihalide monolayers have a trigonal structure with point group symmetry $D_{3d}$. The Cr atoms form a honeycomb lattice with long-range FM order where each Cr (S = 3$/$2) is surrounded by an octahedral environment, coordinated by an edge-sharing octahedra formed by six halide atoms. A sizeable magneto-crystalline anisotropy originated by the spin-orbit coupling (SOC) of the ligands, stabilizes the spin off-plane, with the exception of CrCl$_3$ where the combination of the weak SOC from the Cl atoms and the magnetic dipolar anisotropy energy results in an in-plane ferromagnetic order \cite{shapecrcl,shapecrcl2,shapecrcl3}. Fig. \ref{fig:1} presents the crystal structure of CrX$_3$ and a scheme of the main magnetic exchange interactions between the Cr$^{3+}$ ions, where the interaction between first neighbours, namely J$_1$, takes place through the ligands in a cation-anion-cation super-exchange interaction. Second (J$_2$) and third (J$_3$) order interactions can be explained according to more complex cation-anion-anion-cation pathways. The arrangement of ligands across the honeycomb structure provides a set of Cr-X-Cr and Cr-X-X-Cr angles of close to 90$^{\circ}$  for J$_1$, J$_2$ and to 130$^{\circ}$ in the case of J$_3$ that agrees with the expected FM/AFM behaviour described by the Goodenough-Kanamori-Anderson (GKA) rules. Despite these findings, the microscopic explanation behind the exchange mechanisms has proven to correspond to a more complex picture originated by FM/AFM competitions between $t_{2g}$-$e_{g}$ and $t_{2g}$-$t_{2g}$ spin channels that results in the previous mentioned FM J$_1$, J$_2$ and AFM J$_3$ \cite{katsnelsonorbitally,katsnelsonrelativisticexchange,sorianoenvironmental,spinlatticeresolved}.

To describe the electronic properties of CrX$_3$, we perform fully self-consistent Hubbard-corrected density functional theory (DFT + U) calculations \cite{timrovhp,timrovelmetodo}. The Hubbard U parameter is computed by means of linear response through density functional perturbation theory and this process is repeated in several optimization cycles until convergence of both U and crystal structure are achieved. As a result of this fully free-parameter DFT+U approach, we find that the converged Hubbard U increases as soon as we move down in the periodic table from Cl to I.  In particular, the U values for CrCl$_3$, CrBr$_3$ and CrI$_3$ are 3.96, 4.20 and 4.55 eV, respectively. This evolution naturally follows the expected behaviour of Hubbard U under ligand variation, where the occupations of hybridized transition metals experiment the effect of environment, that enhances the energy to add/remove electrons according to the electronegativity of the ligands (I $>$ Br $>$ Cl) \cite{Uevolution}.

Then, we compute the electronic band structure of CrCl$_3$, CrBr$_3$ and CrI$_3$ for their respective fully optimized structures (Fig. S\ref{fig:bands}). In the octahedral environment of Cr$^{3+}$, the \textit{d} levels split into a lower energy $t_{2g}$ triplet and a higher energy $e_{g}$ doublet, where the three electrons of the magnetic atom occupy the $t_{2g}$ manifold. The calculated electronic structures show band gaps between these two manifolds of 2.00, 1.35 and 0.75 eV for CrCl$_3$, CrBr$_3$ and CrI$_3$, respectively. This decrease from Cl to I indicates the insufficiency of a purely ionic description to characterize the electronic structure of CrX$_3$. The role of the ligand requires to be explained with a more quantitative molecular picture, where the electronic structure is influenced by the hybridization of the ligands, depending on their character, atomic weight and structure \cite{katsnelsonbands,katnelsonbands2}. The orbital resolved density of states, reported in Fig. S\ref{fig:pdos}, shows the most relevant contributing orbitals around the Fermi level. These are mainly \textit{d} orbitals of Cr and \textit{p} orbitals of the halide. One can observe that around the Fermi energy the contribution of \textit{p} orbitals of the halides is extremely important. This indicates a high level of hybridization with the \textit{d} orbitals of Cr. In opposition, the higher energies in the conduction bands are mainly originated by the empty spin-down $t_{2g}$ \textit{d} orbitals of Cr. Furthermore, the more external \textit{p} orbitals in I atoms are weakly bounded to the nucleus resulting in higher energy bands, where in contrast hybridization with Cl atoms arises deeper in energies along the bands. This higher energy hybridization originated by the \textit{p} orbitals at the edges of valence band, result in a narrower band gap in CrI$_3$. \\  

To analyze the magnetic properties of chromium trihalides, we follow an efficient workflow \cite{esterascrsbr} based on purely self-consistent first-principles DFT+U simulations followed by a derived Wannier tight-binding model \cite{wannier90}. This allows to obtain the magnetic exchange interactions (eq. \eqref{eq:1}) in an effective Heisenberg spin Hamiltonian, which is evaluated using the Green's function method as implemented in the TB2J software \cite{tb2j}, without any free parameter. Eq. \eqref{eq:1} presents the spin Hamiltonian, which includes isotropic, symmetric anisotropic and Dzyaloshinskii–Moriya (DM) interactions, as well as a last term considering the single ion anisotropy. This output is then used to obtain the orbital contribution to magnetic exchange, Curie temperature and magnon dispersion (See Computational Methods). This methodology avoids the calculation of exchange interactions using the brute force energy method, and presents a very efficient and easy to automatize route to study magnons in 2D materials. In the case of the CrX$_3$ family, the reduced basis set is formed by the \textit{d} orbitals of Cr and the \textit{p} orbitals of the halide, owing to their major role in stabilizing long-range magnetic order in these materials.  
\begin{equation}
\centering
H = -\sum_{i \neq j}J_{i j} \vec{S}_{i} \cdot \vec{S}_{j}
-\sum_{i \neq j}\vec{D}_{i j} \left(\vec{S}_{i} \times \vec{S}_{j}\right) -\sum_{i}A (\vec{S}_{i}^z)^2
\label{eq:1}
\end{equation}

To provide a microscopic understanding of the magnetic properties, we perform an orbital resolved decomposition of the magnetic exchange interactions. Within this approach, we unveil the contributions of the different orbitals to the magnetic exchange channels and magneto-structural correlations in this class of materials (Fig. S\ref{fig:ccTotal}-S\ref{fig:ciJ3}). As observed in previous works, the most important magnetic exchange interactions correspond to the first neighbour (J$_1$) which are dominated by both $t_{2g}$-$t_{2g}$ (AFM) and $t_{2g}$-$e_{g}$ (FM) orbital channels. Hence, global exchange interactions are originated from the competition between the AFM and FM mechanisms, where the ligands play an important role in the intensity of the exchange interactions. This is illustrated in Fig. \ref{fig:1}, which shows the main interaction channels of CrX$_3$ represented by the Wannier functions of Cr atoms and ligands. A more detailed analysis of the orbital resolved exchange pathways in these three materials can be found in Fig. S\ref{fig:ccTotal}-S\ref{fig:ciJ3}. 

\begin{figure*}[h!]
\includegraphics[scale=0.4]{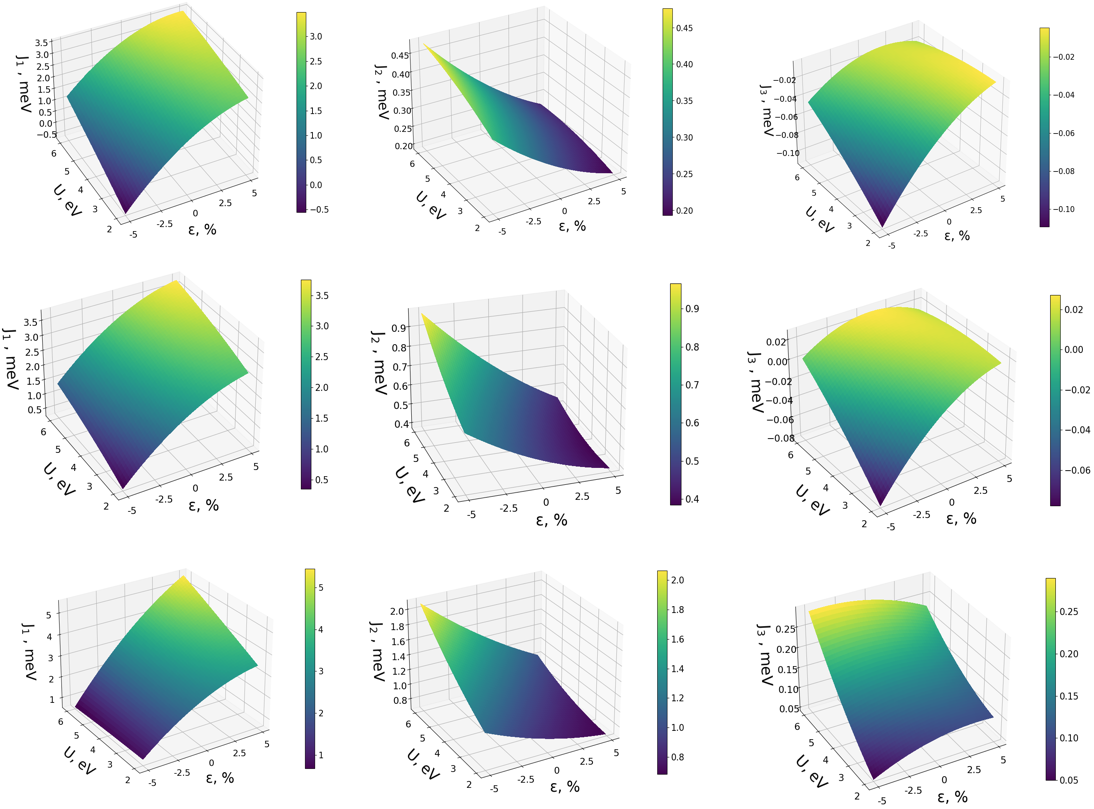}
\caption{ High-density 3D surface plots of isotropic exchange interactions as a function of biaxial strain and Hubbard U in CrCl$_3$ (top), CrBr$_3$ (center) and CrI$_3$(bottom).}
\label{fig:2}
\end{figure*}

Subsequently, we evaluate the computed spin Hamiltonian parameters under applied biaxial strain (see Fig. \ref{fig:2}). For that we distort the in-plane lattice vectors in a range of $\pm$5$\%$ and scan different values of the Hubbard U in a range of 2 - 6 eV. The exchange parameters are condensed in effective equations of the form indicated by eq. \eqref{eq:2} using the least squares method, which are used then to create a dense dataset. 
\begin{equation}
J = \sum_{i=1}^{2} \sum_{j=1}^{2}=a_{i j} U^{i} \epsilon^{j}
\label{eq:2}
\end{equation}

Fig. \ref{fig:2} presents the evolution of exchange interactions for CrX$_3$ where a global enhancement can be observed in all the interactions as long as the Hubbard U is increased. In opposition, the effect of strain is different between J$_2$ and the other interactions. In particular, tensile strain produces a clear enhancement, while in J$_2$, compression is required to increase the intensity of the exchange parameters. This complex behaviour can be successfully understood in terms of the orbital resolved analysis presented in Fig. S\ref{fig:ccTotal}-S\ref{fig:ciJ3}. According to Fig. S\ref{fig:ccTotal}, Fig. S\ref{fig:cbTotal} and Fig. S\ref{fig:ciTotal}, the application of tensile strain to the lattice reduces AFM interactions, thus resulting in a more FM system. The increase of the Hubbard U also supports this enhanced FM character increasing the $t_{2g}$-$e_{g}$ channel. In the particular case of CrI$_3$, an additional element plays a role, the AFM contribution via $d_{z2}$-$d_{z2}$ and $d_{x2y2}$-$d_{x2y2}$ becomes relevant and compete with the FM pathway $d_{z2}$-$d_{x2y2}$, cancelling each other (Fig. S\ref{fig:cri3wf}). As long as the material is elongated, these AFM channels become less important, generating a new source of FM contribution characteristic of CrI$_3$. The increase of Hubbard U enhances globally the different channels that cancel each other, which results in a less important modification of the exchange interactions that can be observed in Fig. \ref{fig:2}, where the evolution of J$_1$ respect to U in CrI$_3$ can be seen as an increment with a very small slope. 

\begin{figure*}[h!]
\includegraphics[scale=0.50]{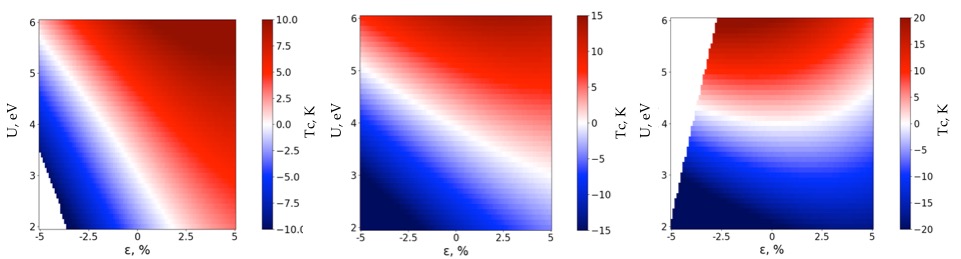}
\caption{ Evolution of Curie temperature under biaxial strain and Hubbard U for (left) CrCl$_3$, (center) CrBr$_3$ and (right) CrI$_3$. }
\label{fig:3}
\end{figure*}

The derived effective equations are used to compute the Curie temperature by solving the quantum Heisenberg model up to third nearest neighbours (see Methods). Using the self-consistent Hubbard U for each material and in absence of strain, we obtain Curie temperature values of 23.0, 55.3 and 94.2 K for CrCl$_3$, CrBr$_3$ and CrI$_3$, respectively, which overestimate the experimental temperatures of 13.0, 34.0 and 45.0 K \cite{amilcar,curiecrbr3,efren}. Nevertheless, they are in good agreement with state-of-the-art calculations based on first principles methods \cite{curie1,curie2}.  Fig. \ref{fig:3} represents the $T_{C}$ evolution in terms of strain and Hubbard U for each CrX$_3$. As one may observe the trends are very similar in the case of CrCl$_3$ and CrBr$_3$, although differences in the amplitude of exchange parameters produce a different derivative of Curie temperature in the maps. On the other hand, CrI$_3$ gives a slightly different trend, which can be rationalized in terms of the dissimilar behaviour of exchange channels (Fig. \ref{fig:2}) and the presence of $e_{g}$-$e_{g}$ interactions, which are characteristic of this material. Regarding the impact of strain on $T_{C}$, our calculations evidence that $T_{C}$ is expected to increase in the three derivatives via stretching up to a 50, 30, 15$\%$ in CrCl$_3$, CrBr$_3$ and CrI$_3$, respectively. In addition, we can observe that when CrCl$_3$ and CrI$_3$ are compressed below $\sim$3$\%$, the maps have regions coloured in white, in which the anisotropy becomes very weak. In these conditions the magnetic anisotropy energy is expected to be fully dominated by shape anisotropy, favouring the in-plane orientation of spins. 

\begin{figure*}[h!]
\includegraphics[scale=0.55]{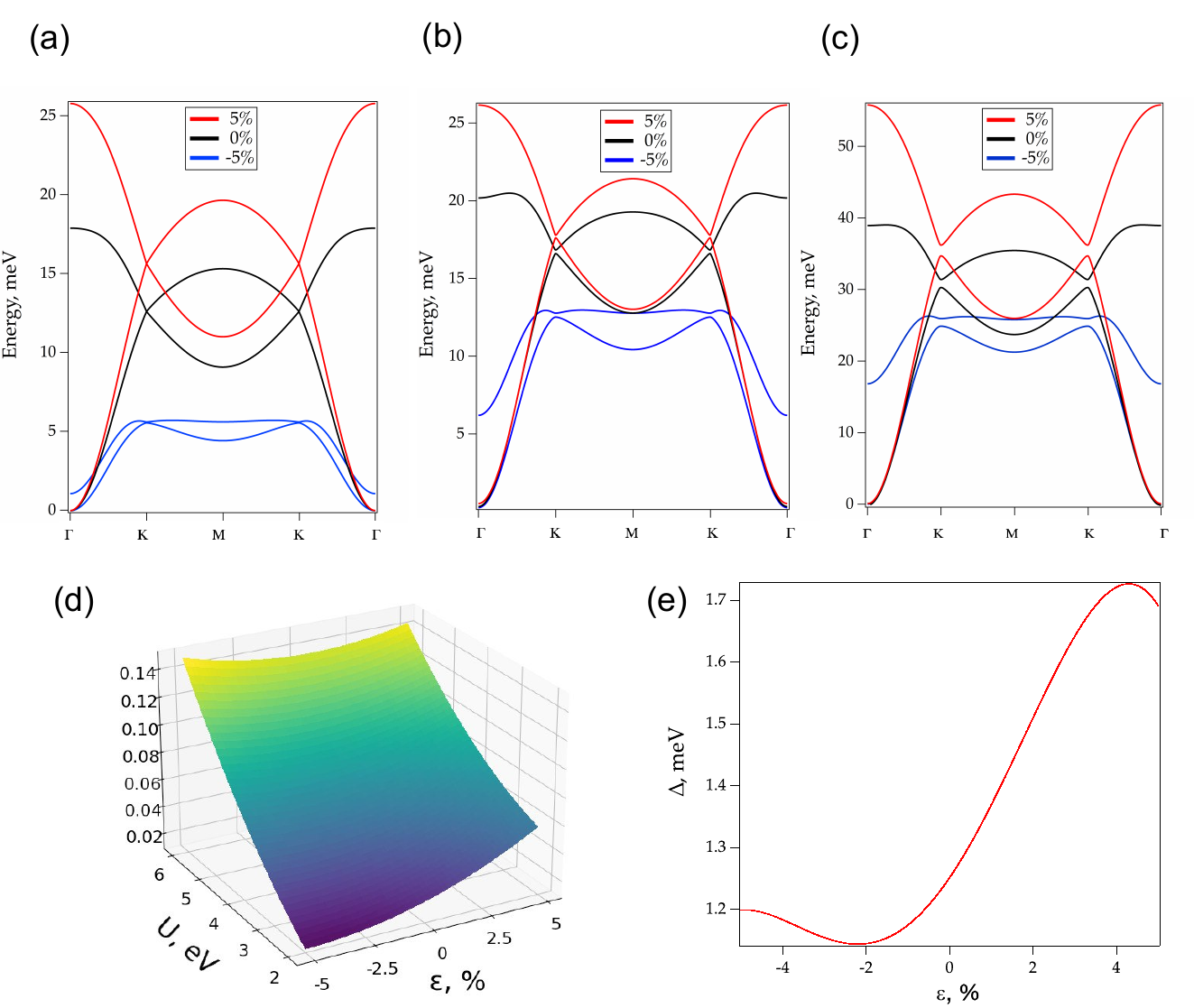}
\caption{ Magnon dispersion of (a) CrCl$_3$, (b) CrBr$_3$ and (c) CrI$_3$, black indicates 0\% strain, and red (blue) 5 (-5)\%. (d) Effective equation representing the z component of the 2nd neighbour DM interaction in CrI$_3$. (e) Evolution of the $\Delta_k$ gap in CrI$_3$. }
\label{fig:4}
\end{figure*}

Afterwards, we calculate the magnon dispersion for each 2D ferromagnet, considering the spin Hamiltonian in eq. \eqref{eq:1} (see methods). In Fig. \ref{fig:4} we can visualize the magnon dispersions of Cr$X_3$, both in equilibrium and under extreme strain conditions ($\pm$5$\%$). For all three chromium trihalides, we obtain an overestimation of the magnon energies respect to the inelastic scattering experiments in the bulk using their optimized single-layer structures, however these findings are in agreement with previous reported calculations \cite{timrovcri3gap,katsnelsonpifiomagnones}. We also simulate the magnon dispersion in the absence of DM interaction (Fig. S\ref{fig:isomagnon}) which shows the typical crossing between acoustic and optical modes forming gapless Dirac cones in the K high-symmetry point of the Brillouin zone in all three compounds. The inclusion of the DM interactions opens a gap $\Delta_k$ (1.26 meV) in CrI$_3$, which is agreement with the values obtained for the bulk in a recently reported DFT study \cite{delugascri3gap,timrovcri3gap}. This opening, results to be purely originated by the DM$_{2z}$ interaction, is absent in CrCl$_3$ and existent but very small in CrBr$_3$ (0.26 meV) due to the small DM interactions arising from the SOC. 

To elucidate the microscopic origin of the DM-driven topological gap, we performed an orbital resolved analysis of DM interactions. It reveals that, in absence of external strain, the symmetry-allowed z component of DM interaction, and thus the topological gap, arise directly (82\%) from the $e_g$-$e_g$ channels of interaction, where the interaction via $d_{x2-y2}$-$d_{z2}$ orbital mechanism results to be dominant. The other (18\%) is due to the competition between different $t_{2g}$-$e_g$ orbitals (Fig. S\ref{fig:resolvedDM}), being the $d_{x2-y2}$ and $d_{xz}$/$d_{yz}$ orbitals the most relevant magnetic exchange pathway. These $t_{2g}$-$e_g$ interactions, tend to cancel each other, given the anisotropic and antisymmetric nature of these interactions. 

In order to analyse the effect of strain in the magnon dispersions we determine the Hubbard U self-consistently for each particular strained structure. Fig. S\ref{fig:banana} illustrates the evolution of Hubbard U under biaxial strain. One can observe that on-site Coulomb repulsion is very sensitive to the compression/elongation of CrX$_3$, giving changes of around 0.5 eV as a function of strain. Then, we simulate the magnon dispersion for the different distorted structures taking into account their corresponding Hubbard U. The magnon dispersion of each material is presented in Fig. \ref{fig:4}, where one can see that the effect of strain produces an important modulation of the magnon modes \cite{milosevicmagnonicsmx3}, (around $\sim$25$\%$ in M high-symmetry point and $\sim$50$\%$ of the maximum frecuency) increasing the dispersive behaviour between $\Gamma$ and K in the case of tensile strain. Furthermore, we find that compressive strain tends to shatter the linear energy dispersion relationship around Dirac cone in K high-symmetry point. Regarding the DM$_{2z}$ interaction evolution under strain, we can observe a drastic enhancement by applying tensile biaxial strain in CrI$_3$ (almost doubled at 5$\%$) as indicated in Fig. \ref{fig:4}. This is because the $t_{2g}$-$e_g$ channels are favoured at the same time that $e_g$-$e_g$ interactions remain constant. As a result, $t_{2g}$-$e_g$ interactions become 40\% of the contribution to DM interaction resulting in the increase of the topological gap. These results contrast with the ones for CrCl$_3$ and CrBr$_3$ where significant changes are not observed (Fig. S\ref{fig:gapsupp}). A possibility to induce a gap opening at the Dirac points of CrBr$_3$ and CrCl$_3$ could be the creation of heterostructures with high SOC materials by proximity effects. This could even be exploited in CrCl$_3$ by achieving a switching between in-plane (Dirac dispersion) and out-of-plane (opened gap) states due to spin-orbit torque effects \cite{prox1,prox2,prox3}.   

\section{Conclusions}
We performed a fully self-consistent first principles analysis of 2D transition metal trihalides motivated by recent findings that suggest the presence of a topological gap in the magnon spectra of bulk CrI$_3$. Our results confirm the existence of a $\Delta_k$ gap opening the Dirac cones of CrI$_3$ down to the 2D limit and discard its presence in CrCl$_3$ and CrBr$_3$. The $\Delta_k$ gap emerges as a direct consequence of DM$_{2z}$ interaction, which supports the presence of topological magnons in 2D CrI$_3$. This points to a competition between DM interaction and interlayer exchange in the bulk, which is rationalized by means of orbital-resolved magnetic exchange calculations. In particular, our microscopic analysis reveals that the $d_{x2-y2}$-$d_{z2}$ orbital mechanism is the source of the DM interaction. We applied biaxial strain to exploit the $t_{2g}$-$e_g$ exchange pathways and increase the robustness of DM interaction, resulting in an increment of the $\Delta_k$ gap. This work illustrates the possibility of tuning topological magnons by strain engineering of DM interaction.

\section{Computational methods}
DFT + U calculations were carried out using the Quantum ESPRESSO package \cite{qe}. The generalized gradient approximation (GGA) and the PBE functional were used to describe the exchange correlation energy \cite{pbe}. We selected standard solid-state US pseudopotentials from the QuantumEspresso database. The electronic wave functions were expanded with well-converged kinetic energy cut-offs for the wave functions and charge density. All the structures were fully optimized using the BFGS algorithm \cite{broyden} until the forces on each atom were smaller than $1\times 10^{-3}$ Ry$/$au, and the energy difference between two consecutive relaxation steps was less than $1\times 10^{-4}$ Ry. The Brillouin zone was sampled by a fine $8\times 8\times 1$ $\Gamma$-centered k-point Monkhorst-Pack mesh for all calculations \cite{monkhorst}. The self consistent Hubbard U and structure were computed until a convergence of 0.01 eV was reached in the Hubbard U. Our DFT + U calculations were followed by Wannier functions calculations using the software Wannier90 \cite{wannier90}. The \textit{d} orbitals of Cr and the \textit{p} orbitals of the ligands, were selected as orbital projectors to maximally localize the Wannier functions. Wannier90 calculations were performed ensuring a correct fit to the electronic band structure and spreads. The orbital resolved analysis was performed after rotating the coordinate system of the crystal to align the metal-ligand bonds direction of the octahedra with the cartesian axes. The quantum Heisenberg model was solved to calculate Curie temperature \cite{belga} and magnon dispersions were derived using the linear spin wave theory \cite{spinw}, combining positive and negative solutions according to Colpa method \cite{colpa}.

\section{Acknowledgements}
The authors acknowledge the financial support from the European Union (ERC-2021-StG-101042680 2D-SMARTiES and FET-OPEN SINFONIA 964396), the Spanish MICINN (2D-HETEROS PID2020-117152RB-100, cofinanced by FEDER, and Excellence Unit “María de Maeztu” CEX2019-000919-M), and the Generalitat Valenciana (grant CDEIGENT/2019/022 and CIDEGENT/2018/004). The computations were performed on the Tirant III cluster of the Servei d’Informàtica of the University of Valencia.

\bibliography{apssamp}
\end{document}